\documentclass[12pt,journal,onecolumn]{IEEEtran}
\usepackage[T1]{fontenc}
\usepackage{amsmath,amssymb,amsthm,mathtools,booktabs}
\usepackage{flafter}
\usepackage[hidelinks,bookmarksnumbered]{hyperref}
\hypersetup{pdftitle={Coding Length and Optimal Power Allocation in Correlated Block Fading},pdfauthor={Haoran Xu and Huazhong Zhao}}
\newtheorem{theorem}{Theorem}[section]
\newtheorem{proposition}[theorem]{Proposition}
\newtheorem{corollary}[theorem]{Corollary}
\newtheorem{lemma}[theorem]{Lemma}
\newcommand{\CN}{\mathcal{CN}}
\newcommand{\Pc}{\mathcal P}

\newcommand{\Ec}{\mathcal E}

\newcommand{\R}{\mathbb R}
\newcommand{\Id}{\mathrm I}
\DeclareMathOperator{\diag}{diag}
\DeclareMathOperator{\tr}{tr}

\newcommand{\doi}[1]{\href{https://doi.org/#1}{doi:\nolinkurl{#1}}}
\begin{document}
\title{Coding Length and Optimal Power Allocation in Correlated Block Fading}
\author{Haoran Xu and Huazhong Zhao%
\thanks{The authors contributed equally. Haoran Xu is the corresponding author.}%
\thanks{H. Xu is with the School of Mathematical Sciences, Soochow University, Suzhou, China (e-mail: 20244007005@stu.suda.edu.cn).}%
\thanks{H. Zhao is with China Telecom Corporation Limited, Suzhou Branch, Suzhou, China (e-mail: 1453785662@qq.com).}}
\maketitle
\begin{abstract}
A codeword spans a fixed number of correlated Gaussian fading blocks.
The receiver knows the gains and the transmitter knows only their law.
At fixed average signal-to-noise ratio, we require the decoding error
probability to decay exponentially with the number of receive antennas.
When does outage predict the achievable rate? If block length grows
faster than the number of antennas, outage gives the limiting optimal
rate. If both grow to infinity at a fixed positive ratio of block length
to array size, outage strictly overestimates the limiting rate. We
determine the critical ratio above which positive rates are possible.
For sufficiently large finite ratios, we find the exact limiting rate
over all codes satisfying a total energy constraint on each codeword.
The leading rate loss is inversely proportional to the ratio.
Its coefficient depends on Gaussian channel dispersion at the most
likely fade causing outage and on how quickly the outage rate decreases
as the required error exponent increases. In the long-block
limit, equal power loses rate on every window of three or more blocks
under stationary first-order autoregressive fading with positive
lag-one correlation. In a different three-block
channel with identical marginal gains, tightening the reliability
requirement first makes it optimal to leave the middle block unused
and later to use it again.
\end{abstract}

\section{Introduction}\label{sec:intro}
A transmitter with no instantaneous channel information must allocate its
energy before it knows which parts of a codeword will fade. If its only
objective is mean mutual information, identical fading marginals make equal
power optimal by concavity. A small error probability poses a different
design problem. In a run of three consecutive blocks, for example, the
middle block can be strongly correlated with both of its neighbors, even
though all three gains have exactly the same distribution. An energy
allocation that performs well on average need not be the best one when a
whole codeword must be decoded reliably.

We consider a single transmit antenna, \(d\) independent receive branches
and \(K\) correlated coherence blocks, with \(L\) uses in each block. The
receiver knows the fading realization. The transmitter knows the temporal
covariance and has one energy budget for the entire codeword. We keep the
mean combined gain fixed as \(d\) increases and ask for the largest rate
at which the average decoding error has exponent at least \(r>0\) per
receive branch. Both the fading and the receiver noise count towards this
error. Which energy allocation attains the rate, and how long must a
coherence block be for the fading alone to determine it?

When \(L/d\) tends to a finite \(\lambda>0\), a less severe fade
together with atypical receiver noise can cause errors more often
than the outage event alone. The limiting rate is strictly below
the long-block outage rate \(R_\infty(r)\). For sufficiently large
\(\lambda\), Theorem~\ref{thm:exact} matches a code with fixed block
energy limits to a converse allowing message-dependent block energies.
Condition \eqref{eq:threshold} depends only on \(r\) and the nominal
signal-to-noise ratio. If \(\overline V\) is the average Gaussian
channel dispersion at the adverse state, the leading rate loss is
\(\overline V/[2K\lambda |R_\infty'(r)|]\).

For shorter blocks the coding and converse bounds can differ.
Pairwise decoding errors still determine which length ratios permit
positive rates: the critical exponent is a log-determinant maximized
over power allocations. If the ratio exceeds its critical value by
\(h\), a construction using nearly orthogonal codewords and a
spherical-code upper bound together give an optimal rate of
\(h^{2+o(1)}\) as \(h\downarrow0\).

When \(L/d\) tends to infinity, the decoding rate reaches the outage
limit. A single adverse covariance of cost \(r\) determines the
optimal powers through water-filling. Equal power is optimal at every
reliability level precisely when each time coordinate has the same
weight in every eigenspace of the fading covariance. Two channels
with the same covariance eigenvalues can therefore call for different
allocations.

\subsection{Related work}
Outage limits the coding rate on slowly varying channels \cite{Ozarow}.
For fixed antenna dimensions, a sphere-packing bound approaches outage
as the code length grows \cite{Foz}. At a fixed target error probability
the quasi-static channel can have zero dispersion \cite{Yang}. A different
regime arises when both antenna dimensions grow with code length:
Hoydis, Couillet and Piantanida \cite{Hoydis} bound the optimal error
near the ergodic rate.

Karadimitrakis, Moustakas and Couillet \cite{Kara} let the transmit
and receive arrays grow in proportion to code length. They derive
the Gallager exponent for rates below the ergodic rate and show that
the random-coding and sphere-packing expressions agree above a
critical rate. The minimizing spectrum already balances the cost of a
fade against conditional decoding error \cite[Remark~1]{Kara}. Their
Gallager bound uses an exponential tilt to account for the codeword
energy constraint. Near the ergodic rate, the dispersion parameter
in their quadratic exponent has a correction of \((2g_0-g_0^2)/\alpha\)
for one fading block. Here \(\alpha\) is code length per transmit
antenna and \(g_0\) is a normalized gain averaged over the typical
limiting spectrum \cite[Sec.~III-C, (30)--(33)]{Kara}.
In our model a single transmit antenna sends across correlated blocks at a
fixed positive error exponent per receive branch. The coefficient
of the rate loss combines the slope of the outage rate at that
exponent with the Gaussian dispersion at the adverse block gains.

For a prescribed gain, the changes in signal amplitude and noise
variance are those of the Gaussian sphere-packing channel
\cite[Sec.~4.1, Example~1, (65)]{Nak}. Our matching equations
determine the adverse gains and the conditional noise together for
each allocation. For sufficiently large length ratios, maximizing
this rate over allocations attains the converse bound for arbitrary
codes, whose block energies may vary with the message. Conditional
information variance appears in the random-coding expansion for a
fixed fading law \cite{Font}. Here each block contributes its Gaussian
dispersion \cite{PPV,PPVparallel} at its adverse gain. Shin and Win
\cite{Shin} derive Gallager bounds for
spatially correlated channels with independent fading blocks.
Kazakopoulos et al.\ \cite{Kaz} study mutual-information outage
as both antenna arrays grow.

Power can be chosen in response to the realization when both ends
know the channel \cite{Caire}. With statistical transmitter information,
previous objectives include ergodic capacity in parallel diversity
channels \cite{Coon}, wideband outage as the number of independent
channels grows \cite{Zhang}, and outage under transmit-covariance
optimization in a multiple-input single-output channel \cite{Abbe}.
Bai et al.\ \cite{Bai} estimate
the probability that the sum of mutual informations of independent
parallel fades falls below a target rate. Our code spans a fixed
number of correlated blocks. Even when the realized gains support the
transmission rate, receiver noise can cause a decoding error and
affect its exponent. A saddle point also appears in
compound matrix channels with prescribed spectral-norm uncertainty
\cite{Loyka}. Here the decoding reliability fixes the size of the
set of adverse channels.

Our related manuscript \cite{XuZhaoPaths} studies the conditional paths
of low continuous-time Gaussian service. Its finite-grid proof of
Theorem~2.3 and covariance contraction (3.4) contain the fixed-power
minimization used here for real temporal covariance and one class of
modes. A deterministic time profile on the active coordinates can be
absorbed by diagonal scaling, and two independent real paths form one
proper complex path. Proposition~7.3 there varies powers on independent
modes under a common time kernel. A separate study of MIMO outage
\cite{XuZhaoStatic} considers fixed equal power and a vanishing target
rate, and determines when the channel may be frozen throughout a
transmission window. The present questions are how to
allocate energy across time and how noise changes the rate of a code
at the prescribed error exponent.

\section{Channel and reliability criterion}\label{sec:model}
Throughout, \(K\) is a fixed positive integer. Let \(C\) be a
positive definite Hermitian \(K\times K\) matrix with diagonal entries
one. The fading vectors \(h_i=(h_{i1},\ldots,h_{iK})^{\mathsf T}\),
\(1\le i\le d\), are independent proper complex Gaussians
\(\CN(0,C)\). Write \(h^{(j)}=(h_{1j},\ldots,h_{dj})^{\mathsf T}\).
In block \(j\), the channel is
\begin{equation}
 Y_{jt}=\sqrt{\rho/d}\,h^{(j)} x_{jt}+Z_{jt},\qquad 1\le t\le L,       \label{eq:channel}
\end{equation}
with \(\rho>0\) and \(Z_{jt}\sim\CN(0,\Id_d)\). The noise vectors are
independent of one another and of the fading. A complex Gaussian of variance one has mean
squared modulus one. The receiver observes all \(h_i\), and the
transmitter has neither feedback nor knowledge of their realizations.
The message is uniform over \(M\) messages and independent of the state.
Each codeword satisfies
\(\sum_{j,t}|x_{jt}|^2\le KL\).
Rates are in nats per complex channel use, and all logarithms are natural.

Given the state, projection onto \(h^{(j)}\) is sufficient for decoding
in block \(j\). The scalar gain squared, apart from \(\rho\), is the
\(j\)th diagonal of the empirical matrix
\[
 X_d=d^{-1}\sum_{i=1}^d h_i h_i^*,\qquad {\mathbb E}X_d=C .
\]
Thus at unit input power the mean combined signal-to-noise ratio is
\(\rho\,{\mathbb E}(X_d)_{jj}=\rho\) for every array size.
Write \(\Pc=\{p\in[0,\infty)^K:\sum_j p_j=K\}\) for the set of
normalized power allocations. Their mutual informations are
\begin{equation}
 S(p,X)=K^{-1}\sum_{j=1}^K\log(1+\rho p_j X_{jj}).                 \label{eq:service}
\end{equation}
The allocations in \(\Pc\) use the full power budget. In a converse, an allocation
with smaller total power can always be enlarged to lie in \(\Pc\).

Fix a sequence of positive integers \(L=L_d\). For \(r>0\), let
\(\mathcal C_{(L_d)}(r)\) be the supremum of \(R\) for which there are
codes satisfying the energy constraint and
\begin{equation}
 \liminf_{d\to\infty}\frac{\log M_d}{KL_d}\ge R,\qquad
 \liminf_{d\to\infty}-\frac{\log P_{e,d}}d\ge r ,                 \label{eq:operational}
\end{equation}
where \(P_{e,d}\) averages over messages, noise and fading.
We hold \(K,C,\rho,r\) fixed as \(d\) and \(L_d\) vary. The covariance
\(C\) describes the same dependence between blocks along each such
sequence. If changing the physical block duration changes that
dependence, the rate must be evaluated with the corresponding \(C\).
The bounds apply to every sequence with \(L_d/d\to\lambda\in(0,\infty)\).
Where they coincide, \(\lambda\) determines the limiting rate.

\section{Reliable rate in long blocks}\label{sec:long}
For a positive definite Hermitian matrix \(X\), define
\begin{equation}
 J_C(X)=\tr(C^{-1}X)-\log\det(C^{-1}X)-K .                           \label{eq:cost}
\end{equation}
The determinant is positive because \(C^{-1}X\) is similar to a positive
definite matrix. Set \(J_C=+\infty\) off the positive definite cone.
We use \(j(y)=y-1-\log y\) for the scalar cost, \(y>0\).
Write \(\Ec_r=\{X\succ0:J_C(X)\le r\}\) and
\(F_r(p)=\min_{X\in\Ec_r} S(p,X)\).

\begin{theorem}[Reliable rate and power allocation]\label{thm:long}
For \(r>0\), if \(L_d/d\to\infty\), every sequence in
\eqref{eq:operational} has the same supremum, namely
\begin{equation}
 R_\infty(r)=\max_{p\in\Pc}\min_{X\in\Ec_r} S(p,X)
             =\min_{X\in\Ec_r}\max_{p\in\Pc} S(p,X).                  \label{eq:game}
\end{equation}
There is a unique saddle pair \((p^*,X^*)\). Write \(x_j=X^*_{jj}\).
There are constants \(\mu>0\) and \(\gamma>0\) such that
\begin{align}
 p^*_j&=\left(\mu-\frac{1}{\rho x_j}\right)_+,\quad \sum_jp^*_j=K,
                                                                        \label{eq:water}\\
 (X^*)^{-1}&=C^{-1}+
 \frac1{K\gamma}\diag\!\left(\frac{\rho p^*_j}{1+\rho p^*_j x_j}\right),
                                                                        \label{eq:worst}
\end{align}
where \(J_C(X^*)=r\).
For any \(p\in\Pc\), the expression
\(\min_{X\in\Ec_r} S(p,X)\) is also the long-block rate attainable with
power at most \(Lp_j\) in each block of every codeword.
\end{theorem}

The covariance \(X^*\) depends on the fading covariance, the signal-to-noise
ratio and the target exponent \(r\).
It describes a collection of states that costs \(r\) in the
large-deviation scale of the receive array. Equation \eqref{eq:water}
uses that one covariance to select a deterministic schedule.

\subsection{Finding the adverse covariance}
The function \(S(p,\cdot)\) is concave, so its minimum over
\(\Ec_r\) requires some care. A scalar variational formula makes both
the global minimum and its uniqueness accessible.

\begin{lemma}\label{lem:inner}
For every \(p\in\Pc\) and \(r>0\) there is exactly one minimizer
\(X_r(p)\) of \(S(p,\cdot)\) over \(\Ec_r\). It has cost \(r\). For a unique
\(\gamma>0\) it is also the unique global minimizer of
\(S(p,X)+\gamma J_C(X)\). Equation \eqref{eq:worst} at \(p\)
together with \(J_C(X)=r\) characterizes this minimizer.
\end{lemma}
\begin{proof}
Put \(a_j=\rho p_j\). For \(u\ge0\), the minimum of
\(t(1+u)-\log t-1\) over \(t>0\) is \(\log(1+u)\) and
is attained at \(t=(1+u)^{-1}\). Apply this identity to each diagonal
entry, and minimize over \(X\) for fixed \(t_j=e^{z_j}\).
The terms depending on \(X\) are linear in \(X\) plus
\(-\gamma\log\det X\). Their unique minimum occurs at
\[
 X=\left(C^{-1}+\diag\{a_j e^{z_j}/(K\gamma)\}\right)^{-1}.
\]
Write \(D_z=\diag\{a_j e^{z_j}/(K\gamma)\}\). Substitution leaves
the following problem on \(\R^K\):
\begin{equation}
 \min_z\ \frac1K\sum_j(e^{z_j}-z_j-1)
 +\gamma\log\det(\Id+CD_z).
                                                                        \label{eq:convex}
\end{equation}
To see why the second term is convex, expand the determinant as
\[
 \sum_{B\subseteq\{1,\ldots,K\}}\det C_{BB}
       \prod_{j\in B}\frac{a_j}{K\gamma}\,
                          \exp\!\left(\sum_{j\in B}z_j\right).
\]
The empty term is one and every coefficient is nonnegative. The
logarithm of this sum is convex by H\"older's inequality. The first
term in \eqref{eq:convex} is strictly convex and tends to infinity
as \(\|z\|\to\infty\). There is thus one minimizing pair. Denote its
matrix by \(X_\gamma\). At the minimum,
\(e^{z_j}=(1+a_j(X_\gamma)_{jj})^{-1}\), giving
\eqref{eq:worst} with \(p\) fixed. Conversely these stationarity
conditions solve the strictly convex problem \eqref{eq:convex}.

The sublevel sets of \(J_C\) are compact in the positive definite
cone. Indeed, each eigenvalue \(y\) of \(C^{-1/2}XC^{-1/2}\)
contributes \(y-1-\log y\), which diverges at zero and infinity.
For \(\gamma\) in a compact positive interval, comparison with \(C\)
bounds \(J_C(X_\gamma)\) uniformly. Compactness and uniqueness then
give continuity in \(\gamma\). The two minimizing inequalities at
\(\gamma_1<\gamma_2\), when added, give
\(J_C(X_{\gamma_2})\le J_C(X_{\gamma_1})\).
Equality would make the two optimizers coincide. Subtracting their
stationarity equations would give \(X=C\), impossible since some
\(a_j>0\). As \(\gamma\to\infty\), comparison with \(C\) gives
\(J_C(X_\gamma)\le S(p,C)/\gamma\). As \(\gamma\downarrow0\),
comparison with \(\epsilon C\), followed by \(\epsilon\downarrow0\),
gives \(S(p,X_\gamma)\to0\), which is impossible on a bounded
cost sublevel. There is exactly one \(\gamma\) of cost \(r\).
The unconstrained minimizing inequality at this \(\gamma\) proves the
constrained result, including uniqueness.
\end{proof}

For each \(X\in\Ec_r\), the diagonal Hessian of \(S(\cdot,X)\) has
entries \(-(\rho X_{jj})^2/[K(1+\rho p_jX_{jj})^2]\).
Compactness gives a common strictly negative upper bound on
\(\Pc\). Taking the lower envelope preserves this strong concavity.
The unique minimizer \(X_r(p)\) varies continuously with \(p\):
every subsequential limit minimizes at the limiting \(p\).
Bounding the two difference quotients for \(F_r\) by evaluating at
the old and new minimizing matrices gives its directional derivative
with gradient
\begin{equation}
 \partial_j F_r(p)=
 \frac{\rho (X_r(p))_{jj}}{K[1+\rho p_j(X_r(p))_{jj}]} .              \label{eq:gradient}
\end{equation}
At the unique maximizing \(p^*\) this gradient is constant on its
positive coordinates and no larger elsewhere. These are exactly the
optimality conditions for maximizing \(S(\cdot,X_r(p^*))\), which
give \eqref{eq:water}. It follows that
\(S(p,X^*)\le S(p^*,X^*)\le S(p^*,X)\) for all admissible \(p,X\).
This proves the equality of the two optimizations in \eqref{eq:game}.
Any other minimizing matrix in the right-hand problem would also
minimize at \(p^*\) and hence would equal \(X^*\).

\subsection{How the decoding error enters}
The empirical matrix \(X_d\) has a large-deviation principle at speed
\(d\) with rate \(J_C\). For completeness, the log moment generating
function of \(hh^*\), evaluated at a Hermitian matrix \(A\), is
\(-\log\det(\Id-C^{1/2}AC^{1/2})\) when the matrix in the determinant
is positive definite. The finite-dimensional Cram\'er theorem
\cite[Th.~2.2.30]{DZ} applies on the real vector space of Hermitian
matrices. The Legendre maximum at a positive definite \(X\) is attained
at \(A=C^{-1}-X^{-1}\) and equals \(J_C(X)\). In a null or negative
direction of \(X\), testing \(A=-t vv^*\) for a unit vector \(v\)
and sending \(t\to\infty\) gives infinite cost.

We give the coding bound for a specified \(p\), since it will also
quantify the effect of a finite length-to-diversity ratio. Draw the
entries of a random codeword independently from \(\CN(0,p_j)\) in
block \(j\) and condition each nonzero block on energy at most
\(Lp_j\). A zero-power block is identically zero. The probability
\(b_L\) of this conditioning event before truncation tends to
\(2^{-|\{j:p_j>0\}|}\) by the central limit theorem in each active
block. Generate the codewords independently from the conditioned law
and use maximum likelihood decoding given the fading.

For \(0<s\le1\) define
\[
 S_s(p,X)=K^{-1}\sum_j\log\{1+\rho p_j X_{jj}/(1+s)\}.
\]
Put \(R_d=(\log M_d)/(KL)\). At a fixed state the expected average
error satisfies
\begin{equation}
 \mathbb E_{\rm code} P_e(X_d)
 \ \le\ \min\{1,\ b_L^{-(1+s)}
                 e^{-KLs[S_s(p,X_d)-R_d]}\}.                          \label{eq:gallager}
\end{equation}
In the random-coding integral of Gallager \cite[Ch.~5]{Gallager},
the input distribution enters through
\[
 \int[\int W(y|x)^{1/(1+s)}\,dP(x)]^{1+s}dy.
\]
Conditioning increases the inner integral by at most \(b_L^{-1}\).
For the original complex Gaussian input the integral factors over
uses and equals \(\prod_j(1+\rho p_j(X_d)_{jj}/(1+s))^{-Ls}\);
this follows by completing the square in the complex Gaussian density.
These two observations give \eqref{eq:gallager}. The receiver's
projection discards independent noise coordinates only.

If \(0<R<F_r(p)\), use codes with \(R_d\to R\), and choose a fixed
\(0<s\le1\) and \(\delta>0\) with
\(R+2\delta+\log(1+s)<F_r(p)\).
The closed set \(S(p,X)\le R+2\delta+\log(1+s)\) has minimum
cost strictly above \(r\), by compactness of the cost sublevels.
Outside this set \(S_s\ge S-\log(1+s)\), so
\eqref{eq:gallager} is at most a constant times \(e^{-KLs\delta}\)
for all sufficiently large \(d\). When \(L/d\to\infty\) this term has infinite exponent at
speed \(d\). Averaging over states and then choosing a deterministic
code proves achievability.

For the converse take any code. Its block energies averaged over
messages and divided by \(L\) give a vector dominated by some
\(p\in\Pc\). Conditional on the state, the Gaussian maximum-entropy
bound, subadditivity over channel uses and concavity of the logarithm
give \(I(U;Y\mid h)\le KL S(p,X_d)\) for the uniform message \(U\),
at each fixed state \(h\). Fano's inequality \cite[Ch.~2]{CoverThomas}
bounds the conditional error from below by
\(1-[S(p,X_d)+\log 2/(KL)]/R_d\).
Let \(R>R_\infty(r)\) and take a matrix
\(X_t=(1-t)X^*+tC\) with \(t>0\) sufficiently small.
Convexity of \(J_C\) gives \(J_C(X_t)\le(1-t)r<r\),
whereas continuity and the saddle inequalities give
\(\max_{\Pc}S(p,X_t)<R\). The latter inequality holds with a
fixed margin on a neighborhood of \(X_t\): there \(S\le R-2\delta\),
while \(R_d\ge R-\delta\) eventually for a code of limiting inferior
rate at least \(R\). The conditional error on that neighborhood is
therefore bounded below by a positive constant. The open-set
large-deviation lower bound puts exponent at most \(J_C(X_t)\) on
the error of any code with limiting inferior rate at least \(R\).
This proves the converse. For fixed block energy bounds the same
argument uses \(X_r(p)\) and \(S(p,\cdot)\) in place of
\(X^*\) and \(\max_{\Pc}S\), completing Theorem~\ref{thm:long}.

The multiplier in \eqref{eq:worst} also gives the slope of the
rate curve:
\begin{equation}
 R_\infty'(r)=-\gamma .                                                \label{eq:slope}
\end{equation}
Indeed, for fixed \(p\), global optimality of the penalized problem
at cost \(r\) gives \(F_{r+h}(p)\ge F_r(p)-\gamma(p,r)h\).
Interchanging \(r\) and \(r+h\) supplies the other bound.
The multiplier is continuous in \((p,r)\) for \(r>0\):
choose two nearby multipliers whose costs straddle \(r\), use compactness
and uniqueness for each penalized problem, and apply strict monotonicity
in the multiplier from Lemma~\ref{lem:inner}. The maximizing allocation
is also continuous in \(r\), by compactness and uniqueness. Applying
the two bounds first at the old maximizing allocation and then at the
new one proves \eqref{eq:slope}.

\section{How much rate does a finite block length cost?}\label{sec:length}
Fix \(r>0\) and let \(L_d/d\to\lambda\in(0,\infty)\).
At the saddle in Theorem~\ref{thm:long}, write \(x_j^*=X^*_{jj}\),
\(g_j=\rho p_j^*x_j^*\) and \(T_j=g_j/(1+g_j)\). The noise
dispersion of a complex Gaussian channel of signal-to-noise ratio \(g\),
in squared nats per use, is \(V(g)=1-(1+g)^{-2}\) \cite{PPV}.
Let \(\overline V=K^{-1}\sum_j V(g_j)\), counting an inactive block
as zero.

\begin{theorem}[Rate loss at a finite length-to-diversity ratio]\label{thm:length}
For every finite \(\lambda>0\),
\(\mathcal C_{(L_d)}(r)<R_\infty(r)\). As \(\lambda\to\infty\),
\begin{equation}
 \mathcal C_{(L_d)}(r)
 =R_\infty(r)-\frac{\overline V}{2K\gamma\lambda}
                +O(\lambda^{-2}).                                     \label{eq:loss}
\end{equation}
Here \(\gamma=-R_\infty'(r)>0\).
The remainder depends only on \(C,K,\rho,r\) and is uniform over
sequences with the given limiting ratio. The rate on the left is
defined by first taking \(d\to\infty\) at fixed \(\lambda\).
\end{theorem}

The numerator in \eqref{eq:loss} uses the gains at the rare channel
state. Using the typical gains \(C_{jj}=1\) would generally give a
different coefficient. To derive it we compare a code with two nearby
Gaussian channels. On the coding side a change in input variance
accounts for the energy of each codeword. On the converse side the
receiver sees a slightly weaker signal and slightly greater noise.

\subsection{A coding bound that retains the energy constraint}
Fix \(p\in\Pc\). In block \(j\) draw independent Gaussian symbols
of variance \(p_j/(1-u_j)\), then condition on total block energy at
most \(Lp_j\), with \(0\le u_j<1\). Use zero symbols and \(u_j=0\)
on an inactive block. Before conditioning, the normalized energy of
an active block is the mean of \(L\) independent unit exponentials.
The probability of passing the energy test has negative log divided
by \(L\) tending to \(j(1-u_j)=-u_j-\log(1-u_j)\). This follows by
tilting the exponentials to mean \(1-u_j\), or by the scalar
Cram\'er theorem. At \(u_j=0\) the probability tends to \(1/2\).
For \(0<s\le1\), set
\[
 S_{s,u}(p,X)=\frac1K\sum_j
       \log\left(1+\frac{\rho p_jX_{jj}}{(1-u_j)(1+s)}\right).
\]
Use the Gaussian integral in \eqref{eq:gallager}, including the
conditioning factor to power \(1+s\), and average over the fading.
The resulting achievable rate is at least
\begin{equation}
 \begin{aligned}
 &\inf_{X\succ0}\left\{S_{s,u}(p,X)+
                  \frac{J_C(X)-r}{K\lambda s}\right\}\\
 &\hspace{2em}+\frac{1+s}{Ks}\sum_j[u_j+\log(1-u_j)].
 \end{aligned}                                                         \label{eq:ach}
\end{equation}
Indeed the expectation of the untruncated Gallager bound involves
\(\mathbb E\exp[-KL_ds S_{s,u}(p,X_d)]\). The function
\(-S_{s,u}\) is continuous and bounded above on the positive
semidefinite cone. The large-deviation upper bound for integrals
\cite[Th.~4.3.1]{DZ} therefore applies. Replacing \(L_d/d\) by its
limit is uniform on cost sublevels, and exponential tightness controls
the rest since the integrand is at most one. Solving the resulting
exponent bound for the rate gives \eqref{eq:ach}.

Set \(p=p^*\), \(s=1/(K\lambda\gamma)\) and \(u_j=sT_j\), which
are admissible for large \(\lambda\). As \(s\downarrow0\) the
unperturbed objective under the infimum is
\(S(p^*,X)+\gamma[J_C(X)-r]\). Its unique minimum is \(R_\infty(r)\).
We record why taking its first-order change at \(X^*\) has an
\(O(s^2)\) remainder. In the joint minimization over \(X,z\) in
Lemma~\ref{lem:inner}, the Hessian block in \(X\) is the positive
quadratic form \(\gamma\tr(X^{-1}H X^{-1}H)\) for \(H=H^*\ne0\).
Its Schur complement is the Hessian of \eqref{eq:convex}, which is
positive definite because of \(\sum_j e^{z_j}/K\). The full Hessian
is positive definite. Eliminating \(z\) instead proves positive
definiteness of the Hessian of \(S+\gamma J_C\) at \(X^*\).
All minimizers of the perturbed problem stay in a fixed compact cost
sublevel since \(S_{s,u}\ge0\); they converge to \(X^*\) by uniqueness.
The implicit function theorem then makes the displacement \(O(s)\).
Taylor expansion at \(X^*\) gives the asserted remainder.

At that matrix the derivative of the \(j\)th logarithm in
\(S_{s,u}\) is \(T_j(T_j-1)\). The last term of \eqref{eq:ach}
contributes \(-s\sum_jT_j^2/(2K)+O(s^2)\). Since
\(2T_j-T_j^2=V(g_j)\), their sum yields the lower bound in
\eqref{eq:loss}. These steps also cover \(p_j^*=0\).

\subsection{A converse allowing different block energies for each message}
Let \(\overline{\Pc}=\{p\ge0:\sum_jp_j\le K\}\). Any achievable
positive rate \(R\) has a profile \(p\in\overline{\Pc}\) with the
following property. Let \(b_j\in[0,1)\) denote a fractional reduction
in signal amplitude and \(w_j\ge1\) a noise variance. For any
\(X\succ0\) put
\begin{align}
 D(p,X;b,w)&=J_C(X)+\lambda\sum_j
             \bigl[j(w_j)+\rho p_jX_{jj} b_j^2\bigr],                 \label{eq:budget}\\
 S(p,X;b,w)&=\frac1K\sum_j\log\left(1+
             \frac{\rho p_jX_{jj}(1-b_j)^2}{w_j}\right).
                                                                            \label{eq:localrate}
\end{align}
Then
\begin{equation}
 D(p,X;b,w)<r\quad\Longrightarrow\quad R\le S(p,X;b,w).
                                                                        \label{eq:jointconverse}
\end{equation}

To find \(p\), divide the possible block energy vectors of the
codewords, normalized by \(L_d\), into cubes of side \(1/d\).
At most \((Kd+1)^K\) cubes are needed. Keep the most populated
cube. The rate loss is \(o(1)\), and the average error on this
subcode is at most \((Kd+1)^K P_{e,d}\), so the required exponent
is preserved. Along a subsequence every message in the cube has
block energies per use \(p_j+o(1)\), uniformly over messages, for a
single \(p\in\overline{\Pc}\).

Suppose that \(D<r\) and \(R>S(p,X;b,w)\). Under an alternative law,
condition the fading on \(X_d\) belonging to an open ball of radius
\(\delta\) about \(X\). For each channel use multiply the conditional
signal amplitude by \(1-b_j\) and change
the noise variance to \(w_j\) in the known signal direction only.
The orthogonal noise coordinates retain their original law. Fano's
inequality and the Gaussian entropy bound on this ball give a positive
lower bound on the alternative error, for sufficiently small \(\delta\).
The upper bound on mutual information per use differs from
\eqref{eq:localrate} by \(O(\delta)+o(1)\).

The fading contribution to the log likelihood of the alternative
relative to the original law is minus the logarithm of the probability
of the ball. Its limiting superior after division by \(d\) is at
most \(J_C(X)\) by the open-set large-deviation bound. For the
noise contribution, write \(\mu\) for the nominal projected signal
and \(W\sim\CN(0,w_j)\) for the alternative noise. Each
log likelihood is
\[
 -\log w_j+(1-1/w_j)|W|^2-2b_j\Re(\overline W\mu)+b_j^2|\mu|^2 .
\]
Conditional on message and state it has mean \(j(w_j)+b_j^2|\mu|^2\)
and variance \((w_j-1)^2+2w_j b_j^2|\mu|^2\). The variables are
independent over uses. On the ball,
\(\sum_t|\mu_{jt}|^2/L_d=\rho(X_d)_{jj}[p_j+o(1)]\) uniformly.
The total conditional variance divided by \(d^2\) tends to zero.
Thus, with alternative probability tending to one, the normalized
joint log likelihood is at most \(D+O(\delta)+o(1)\), up to any
fixed positive tolerance. Intersect this event with the error event
and change back to the original law. Taking \(\delta\) and the
tolerance small enough forces the error exponent below \(r\).
This proves \eqref{eq:jointconverse}.

\subsection{When the two bounds agree exactly}
The comparison channels can be chosen to make \eqref{eq:ach} and
\eqref{eq:jointconverse} agree before expanding in \(1/\lambda\).

\begin{theorem}[A finite-ratio coding formula]\label{thm:exact}
Fix \(p\in\Pc\). For \(\Gamma>0,\ X\succ0\), write
\(s=1/(K\lambda\Gamma)\), \(g_j(p,X)=\rho p_jX_{jj}\).
For \(g=g_j(p,X)\) let \(\xi_j\) be the positive root of
\begin{equation}
 g\xi^2+(1+s-g)\xi-1=0,\qquad w_j=(1+s)\xi_j.                       \label{eq:matchscalar}
\end{equation}
Consider the equations
\begin{align}
 X^{-1}&=C^{-1}+\frac1{K\Gamma}
     \diag\!\left(\frac{\rho p_j\xi_j}{1+g_j(p,X)\xi_j}\right),       \label{eq:matchX}\\
 r&=J_C(X)+\lambda\sum_j
          [j(w_j)+g_j(p,X)(1-\xi_j)^2].                              \label{eq:matchcost}
\end{align}
Whenever these equations have a solution with \(0<s\le1\), the
operational rate under the individual block energy limits \(Lp_j\)
is exactly
\begin{equation}
 G_\lambda(p)=\frac1K\sum_j
       \log\left(1+\frac{g_j(p,X)\xi_j^2}{w_j}\right).                \label{eq:exactrate}
\end{equation}
There exists a finite \(\Lambda(C,K,\rho,r)\) such that all profiles
have such a solution for every \(\lambda\ge\Lambda\), and
\(\mathcal C_{(L_d)}(r)=\max_{p\in\Pc}G_\lambda(p)\) under the single
total energy constraint. The formula holds for every sequence with
limiting ratio \(\lambda\). An explicit sufficient ratio is given
in \eqref{eq:threshold}.
\end{theorem}
\begin{proof}
The polynomial in \eqref{eq:matchscalar} has one positive root
in \([1/(1+s),1)\), including at \(g=0\). Suppress the block
subscript and take \(u=1-1/w\) in the code and \(b=1-\xi\) in
the converse. From \((1-\xi)(1+g\xi)=s\xi\) we obtain
\begin{equation}
 \frac{1+g\xi}{1+g\xi^2/w}=w,\qquad
 g(1-\xi)^2=(1+s)u-(w-1).                                             \label{eq:matchid}
\end{equation}
In \eqref{eq:ach}, hold these \(u_j\) fixed while minimizing over
the fading matrix. Its stationarity equation is \eqref{eq:matchX},
since \(1/[(1-u_j)(1+s)]=\xi_j\). The proof of Lemma~\ref{lem:inner},
with coefficients \(\rho p_j\xi_j\), gives global minimality.
Insert the budget \eqref{eq:matchcost}. By the second identity in
\eqref{eq:matchid} the \(j\)th contribution from the cost and the
energy conditioning term simplifies to \(-\log(w_j)/K\). The first
identity gives the achievable rate \eqref{eq:exactrate}.
In the converse the same parameters give that rate. To meet the
strict budget, shrink \(b_j,w_j-1\) slightly toward zero and pass to
the limit. Some block has \(g_j>0\), hence \(w_j>1\), so its cost
strictly decreases. Both cost and rate are nondecreasing in the
message energy profile. The test therefore covers every subcode
profile dominated by \(p\).

For existence put \(\epsilon=1/\lambda\) and
\(s=\epsilon/(K\Gamma)\). At \(s=0\) the root \(\xi=1\) has
polynomial derivative \(1+g>0\), so it is smooth in \(s,g\) near
this point. Both \(w-1\) and \(1-\xi\) are \(O(\epsilon)\).
The noise cost in \eqref{eq:matchcost}, including its factor
\(\epsilon^{-1}\), extends smoothly to \(\epsilon=0\).
Equations \eqref{eq:matchX}--\eqref{eq:matchcost} reduce there to the
constrained stationarity equations at \(X_r(p),\gamma(p,r)\).
Writing the first equation in gradient form, the derivative in
\((X,\Gamma)\) has the bordered Hessian
\[
 \begin{pmatrix}
 \nabla^2_X(S+\gamma J_C)&\nabla J_C\\
 (\nabla J_C)^{\mathsf T}&0
 \end{pmatrix}
\]
in real coordinates for Hermitian matrices. Its leading block is
positive definite by the Schur-complement argument above and
\(\nabla J_C\ne0\) since \(J_C=r>0\). The derivative is invertible.
The implicit function theorem gives a smooth solution near each
\((p,0)\), including boundary allocations: the equations extend
to an open neighborhood in the real \(p\) coordinates. Finitely
many such neighborhoods cover the simplex, giving a common positive
range of \(\epsilon\). On this branch \(X\succ0,\Gamma>0\) and \(s\le1\).
Overlapping neighborhoods give the same rate by the matching just
proved. For a code with unrestricted block energies take its subcode
profile from \eqref{eq:jointconverse} and enlarge it to \(p\in\Pc\).
Its rate is at most \(G_\lambda(p)\). Conversely the individual
limits at each \(p\in\Pc\) respect the total budget, proving the
maximization formula.
\end{proof}

There is also an explicit sufficient length. Let \(a_r\in(0,1)\)
solve \(j(a_r)=r\). For \(g>0\) take the root \(\xi\) of
\eqref{eq:matchscalar} at \(s=1\) and put
\[
 c(g)=2-\xi^{-1}-\log(2\xi).
\]
Then the exact maximization formula holds whenever
\begin{equation}
 \lambda> r/c(\rho a_r).                                               \label{eq:threshold}
\end{equation}
To prove this, fix \(p,\lambda\), choose \(s\in(0,1]\), and set
\(\Gamma=1/(K\lambda s)\). In \eqref{eq:ach} maximize the lower
bound over \(v_j=1-u_j\in(0,1]\). Before minimizing over \(X\), the
terms involving one \(v\), multiplied by \(Ks\), are
\[
 s\log[(1+s)v+g]+\log v+(1+s)(1-v)-s\log(1+s).
\]
These terms are strictly concave in \(v\), with a common negative
upper bound on their second derivative. Their lower envelope over
\(X\) is strictly concave as well, and tends to minus infinity at a
zero coordinate, by evaluating at \(X=C\). It has one maximizer.
For fixed \(v\) the inner adverse covariance is unique by
Lemma~\ref{lem:inner}. The derivative of the envelope in \(v\)
is therefore the derivative at that covariance. The scalar maximizing
condition gives \(v=1/[(1+s)\xi]\) by \eqref{eq:matchscalar}, also
at \(g=0\) where \(v=1\). Thus \eqref{eq:matchX} holds. This pair
depends continuously on \(s>0\), by uniqueness and compactness on
closed subintervals of \((0,1]\). Denote its joint cost, the
right side of \eqref{eq:matchcost}, by \(D(s)\).
As \(s\downarrow0\), \(1\le w\le1+s\) and \(0\le 1-\xi\le s\).
Comparison with \(C\) in the inner minimization gives
\(J_C(X)\le S_{s,u}(p,C)/\Gamma\to0\). The gains are consequently
bounded and \(D(s)\to0\).

If \(D(1)\le r\), then \(J_C(X)\le r\). Each eigenvalue \(y\) of
\(C^{-1/2}XC^{-1/2}\) satisfies \(j(y)\le r\), so \(X\succeq a_r C\).
Some \(p_j\ge1\), hence \(g_j\ge\rho a_r\). At \(s=1\),
\eqref{eq:matchid} gives \(j(w)+g(1-\xi)^2=c(g)\). The root \(\xi\)
increases with \(g\), and \(c\) increases from zero because
\(d(2-\xi^{-1}-\log(2\xi))/d\xi=(1-\xi)/\xi^2>0\).
Thus \(D(1)\ge\lambda c(\rho a_r)>r\), a contradiction.
Continuity now supplies a solution \(D(s)=r\) for every \(p\).
For a sequence of profiles approaching the supremum, the corresponding
matrices lie in \(\Ec_r\). The preceding estimate at small \(s\) is
uniform in \(p\): \(S_{s,u}(p,C)\le\log(1+\rho K)\). Hence
\(r>0\) keeps \(s\) away from zero. A convergent subsequence of
profiles, matrices and parameters gives a matched solution attaining
the maximum.
Bound \eqref{eq:threshold} is sufficient; an individual profile
can satisfy the matching equations at shorter lengths.

\subsection{The strict loss and its coefficient}
First, \eqref{eq:jointconverse} gives a strict loss at every finite
\(\lambda\). Take \(X_t=(1-t)X^*+tC\), \(b_j=0\), \(w_j=1+v\),
and \(t=K\lambda v^2/r\). For small \(v>0\), convexity and
\(j(1+v)\le v^2/2\) give \(D<r\) for every profile.
Maximizing allocations for \eqref{eq:localrate} exhaust the available
energy, so some \(p_j\ge1\). Put \(x_{\min}=\min(1,\min_j x_j^*)>0\).
Increasing the variance from 1 to \(1+v\le2\) lowers the rate by at
least \(v\rho x_{\min}/[2K(2+\rho x_{\min})]\), whereas moving the
covariance from \(X^*\) to \(X_t\) costs at most
\(\rho t\max_j|1-x_j^*|\) in rate. For sufficiently small \(v\),
the maximum in \eqref{eq:localrate} is strictly below \(R_\infty\).

For the coefficient use the smooth branch in Theorem~\ref{thm:exact}
and evaluate the matched rate by \eqref{eq:ach}. At \(\epsilon=0\),
changes in \(X\) and \(\Gamma\) have zero first derivative in
\(S+\Gamma(J_C-r)\), by constrained stationarity.
Differentiating \eqref{eq:matchscalar} gives
\(u_j=sT_j(p)+O(s^2)\), where
\(g_j(p)=\rho p_j[X_r(p)]_{jj}\) and \(T_j(p)=g_j(p)/(1+g_j(p))\).
The calculation following \eqref{eq:ach} gives the first derivative
of the rate at each \(p\):
\begin{equation}
 \left.\frac{\partial}{\partial\epsilon}G_{1/\epsilon}(p)
       \right|_{\epsilon=0^+}=-B(p),\qquad
 B(p)=\frac{\sum_j V(g_j(p))}{2K^2\gamma(p,r)} .                   \label{eq:expandQ}
\end{equation}
Here \(G_{1/\epsilon}(p)\) extends to \(F_r(p)\) at \(\epsilon=0\).
The smooth solutions in Theorem~\ref{thm:exact} extend across the
faces of \(\Pc\), including allocations with inactive blocks.
Choose finitely many closed neighborhoods inside these solution
domains whose interiors cover \(\Pc\times\{0\}\). The second
\(\epsilon\)-derivative of the rate is bounded on each. Taylor's
theorem therefore gives constants \(A<\infty,\epsilon_0>0\),
depending only on \(C,K,\rho,r\), such that for \(0<\epsilon\le\epsilon_0\),
\begin{equation}
 \sup_{p\in\Pc}
 \left|G_{1/\epsilon}(p)-F_r(p)+\epsilon B(p)\right|
 \le A\epsilon^2.                                                    \label{eq:uniformprofile}
\end{equation}
This bounds the remainder in the limiting rate at each fixed ratio.
The same smoothness makes \(B\) Lipschitz on \(\Pc\). Strong concavity
gives \(F_r(p)\le R_\infty-c\|p-p^*\|^2\) for some \(c>0\).
With \(z=\|p-p^*\|\), the possible increase in \(F_r-\epsilon B\)
over its value at \(p^*\) is at most
\(-c z^2+\epsilon\,\mathrm{Lip}(B)z\le
 \epsilon^2\mathrm{Lip}(B)^2/(4c)\).
Evaluating \eqref{eq:uniformprofile} at \(p^*\) for the lower bound
and taking the supremum for the upper bound gives \eqref{eq:loss}.

\subsection{The boundary for a positive coding rate}\label{sec:positive}
The matching equations give the rate for sufficiently long coherence
blocks. At the other end of the range we can decide, for every length
ratio, whether any positive rate is possible. Define
\begin{equation}
 r_{\rm pos}(\lambda)=
 \max_{p\in\Pc}\log\det\!\left(\Id+
              \frac{\lambda\rho}{2} C\diag(p)\right).                  \label{eq:positive}
\end{equation}

\begin{theorem}[Positive-rate boundary]\label{thm:positive}
At every fixed \(\lambda\in(0,\infty)\) and \(r>0\),
\[
  \mathcal C_{(L_d)}(r)>0\quad\Longleftrightarrow\quad
                r<r_{\rm pos}(\lambda).
\]
More precisely, every achievable \(R>0\) satisfies
\(r\le r_{\rm pos}(\lambda e^{-R})\). If \(\lambda_{\rm pos}(r)\)
is the unique root of \(r_{\rm pos}(\lambda)=r\), this gives the
upper bound \(\mathcal C_{(L_d)}(r)\le
[\log(\lambda/\lambda_{\rm pos}(r))]_+\) at every ratio.
\end{theorem}
\begin{proof}
For a pair of codewords let \(D_j\) be their squared Euclidean distance
in block \(j\). Given the channel, the minimum average error in testing
the two equiprobable messages is \(Q(\sqrt{a/2})\), where
\(a=\rho\sum_j (X_d)_{jj}D_j\) and \(Q\) is the standard normal upper
tail. The factor \(1/2\) inside the square root uses the complex noise
normalization in \eqref{eq:channel}. For every \(v\ge0\) the fading
average of the exponential distance is
\begin{equation}
 \mathbb E e^{-v a/4}=
 \det\!\left(\Id+\frac{v\rho}{4d} C\diag(D_1,\ldots,D_K)\right)^{-d}.
                                                                         \label{eq:pair}
\end{equation}
This follows by integrating the complex Gaussian density of each
receive branch. Determinant expressions for pairwise errors over
fading are classical in space--time coding \cite{Tarokh,Dog}. We use
the distribution of distances in a positive-rate code in addition to
the pairwise bound.

Take \(p\) attaining \eqref{eq:positive} and write \(n=KL_d\).
On the unit sphere in \(\mathbb C^{L_d}\), the directions whose inner
product with a fixed unit vector has modulus greater than \(\delta\in(0,1)\)
occupy a fraction \((1-\delta^2)^{L_d-1}\) of the sphere. This follows
by normalizing a vector of independent complex Gaussians. After choosing
a direction, exclude this set and choose the next from what remains.
For \(KR<-\log(1-\delta^2)\) we can thus choose \(\lfloor e^{nR}\rfloor\)
directions for all sufficiently large \(d\). In block \(j\) send the
direction of the message scaled by \(\sqrt{L_dp_j}\). Every pair then
satisfies \(D_j\ge2L_dp_j(1-\delta)\) in every block.
Using \(Q(\sqrt{a/2})\le e^{-a/4}/2\), \eqref{eq:pair} and a
union bound over competing messages gives error exponent at least
\[
 \log\det\!\left(\Id+\frac{\lambda\rho(1-\delta)}2C\diag(p)\right)
       -K\lambda R .
\]
For \(r<r_{\rm pos}(\lambda)\), first choose \(\delta\) small enough,
then \(R>0\) small enough, to exceed \(r\).

For the converse, use the spherical geometry of Gaussian codes
\cite{Shannon}. Lift each complex codeword \(x\) to
\((\Re x,\Im x,\sqrt{n-\|x\|^2})\) on the sphere of radius
\(\sqrt n\) in \(\R^{2n+1}\). This can only increase distances.
Fix \(0<R_0<R\) and take \(t>0\) with \(1-t^2=e^{-R_0}\).
For a uniform unit vector \(U\), let the list consist of messages
whose lifted codewords have projection at least \(t\sqrt n\) on \(U\).
Each message is listed with the same probability
\(a_n=\exp[-nR_0+o(n)]\), since the first coordinate on the unit
sphere has density proportional to \((1-z^2)^{n-1}\).

Keep the messages whose individual error probability, averaged over
fading and noise, is at most \(2P_{e,d}\). At least \(M/2\) messages
remain. Some axis therefore lists \(N\ge Ma_n/2\) retained messages,
and \(N\to\infty\) since \(R_0<R\). The mean lifted codeword in this
list has projection at least \(t\sqrt n\) on its axis. Half the average
squared distance over distinct pairs in the list equals
\[
 \frac{N}{N-1}\bigl(n-\|\text{mean lifted codeword}\|^2\bigr)
 \le (1+o(1))n(1-t^2).
\]
Select a pair no farther apart than this average. Its original block
distances obey \(\sum_jD_j/(2L_d)\le Ke^{-R_0}(1+o(1))\). The
optimal binary error, averaged over fading, is bounded above by
the mean of the two original message errors and hence by
\(2P_{e,d}\). For each \(\zeta>0\) there is \(c_\zeta>0\) such that
\(Q(\sqrt{a/2})\ge c_\zeta e^{-(1+\zeta)a/4}\) for all \(a\ge0\),
by the normal tail asymptotic and continuity. Averaging this lower
bound for the selected pair by \eqref{eq:pair} gives
\[
 \limsup_d -d^{-1}\log P_{e,d}\le
 r_{\rm pos}\bigl((1+\zeta)\lambda e^{-R_0}\bigr).
\]
The determinant increases in nonnegative powers. The maximum in
\eqref{eq:positive} is continuous and strictly increasing from
zero to infinity, by compactness and strict increase at every
full-budget allocation. Let \(\zeta\downarrow0\) and \(R_0\uparrow R\).
This proves the upper bound and excludes positive rate at equality.
\end{proof}

For one block this gives a necessary and sufficient condition for
positive rate, \(\lambda>2(e^r-1)/\rho\). With a fixed two-message
alphabet, two antipodal codewords give \(D_1=4L_d\) and exponent
\(\log(1+\lambda\rho)\). Hence allowing only finitely many messages
would answer a different reliability question.

\begin{corollary}[Rate near the positive-rate boundary]\label{cor:opening}
Fix \(K,C,\rho,r\), set \(\lambda_0=\lambda_{\rm pos}(r)\) and
\(\epsilon=1-\lambda_0/\lambda\). As \(\lambda\downarrow\lambda_0\),
every sequence with \(L_d/d\to\lambda\) satisfies
\[
 (1-o(1))\frac{\epsilon^2}{K}\ \le\ \mathcal C_{(L_d)}(r)
 \ \le\ \epsilon^2\log(1/\epsilon)+O(\epsilon^2).
\]
In particular the ratio of \(\log\mathcal C_{(L_d)}(r)\) to
\(\log(\lambda-\lambda_0)\) tends to two.
\end{corollary}
\begin{proof}
Use the code in the proof of Theorem~\ref{thm:positive}, whose
directions can be chosen whenever \(KR<-\log(1-\delta^2)\).
Choose \(p\) maximizing
\eqref{eq:positive} at \(\lambda_0\),
\(\delta=\epsilon-\epsilon^{3/2}\) and \(R=\delta^2/K\).
The determinant logarithm at this fixed \(p\) has positive derivative
at \(\lambda_0\). Since \(\lambda(1-\delta)=\lambda_0+
\lambda\epsilon^{3/2}\), its increase exceeds the union-bound cost
\(K\lambda R=O(\epsilon^2)\) for small \(\epsilon\).

For the converse keep the messages with individual error at most
\(2P_{e,d}\). For any retained pair put \(q=\sum_jD_j/(2KL_d)\).
Its binary error is at least
\(c_\zeta\exp[-d\,r_{\rm pos}((1+\zeta)(L_d/d)q)]\)
by \eqref{eq:pair}, and at most \(2P_{e,d}\).
Letting \(d\to\infty\) and then \(\zeta\downarrow0\) gives
\(\liminf_d\min q\ge\lambda_0/\lambda\).
The lifted unit vectors from Theorem~\ref{thm:positive} therefore have
pairwise real inner products at most \(\epsilon+o_d(1)\).
Apply the Kabatiansky--Levenshtein bound
\cite[Th.~4]{KL} at any fixed cosine above \(\epsilon\), then decrease
it to \(\epsilon\). There are \(2KL_d+1\) real dimensions, so
\(R\le2[(1+u)\log(1+u)-u\log u]\), where
\(u=((1-\epsilon^2)^{-1/2}-1)/2=\epsilon^2/4+O(\epsilon^4)\).
Expansion gives the stated upper bound.
\end{proof}

Thus at or below \(\lambda_{\rm pos}(r)\) the optimal rate is zero.
Just above it, Corollary~\ref{cor:opening} determines the logarithmic
order of the rate, with a logarithmic factor between its two bounds.
At larger ratios, \eqref{eq:ach} and \eqref{eq:jointconverse} give
coding and converse bounds. For a given allocation they agree when
\eqref{eq:matchscalar}--\eqref{eq:matchcost} have a solution with
\(0<s\le1\). Condition \eqref{eq:threshold} guarantees such a solution
for every allocation, so maximizing \eqref{eq:exactrate} gives the
rate under the total energy constraint. At intermediate ratios where
the bounds have yet to be matched, the exact positive rate remains
open in general. As \(\lambda\to\infty\), \eqref{eq:loss} gives the
leading correction to the outage rate.

\subsection{An example with non-Gaussian fading amplitudes}
The noise term in \eqref{eq:loss} also arises when the channel
amplitudes have a different law. Suppose the squared fading amplitudes
are independent across branches and blocks, each gamma distributed
with shape and rate \(m>0\). As in Section~\ref{sec:model}, the
receiver knows the full fading realization and the transmitter knows
only its law.
The normalized gains \(y_j=d^{-1}\sum_i |h_{ij}|^2\) have rate
\(J(y)=m\sum_j j(y_j)\), by the scalar Cram\'er theorem and independence.
Let \(y\in(0,1)\) solve \(Kmj(y)=r\). Then the long-block rate is
\(\log(1+\rho y)\), and the loss at large \(\lambda\) is
\begin{equation}
 \frac{m(1-y)(2+\rho y)}{2(1+\rho y)\lambda}
       +O(\lambda^{-2}).                                               \label{eq:nak}
\end{equation}
Here is a direct verification of the hypotheses used above. The power
objective is concave and invariant under permutations, so equal power
maximizes it. Put \(\gamma=\rho y/[Km(1+\rho y)(1-y)]\).
For each gain \(z>0\), the derivative of
\(\log(1+\rho z)/K+\gamma m j(z)\) has the sign of
\[
 K\gamma m\rho z^2+[\rho+K\gamma m(1-\rho)]z-K\gamma m .
\]
Its unique positive root is \(y\), and the derivative changes sign
there with strictly positive slope. Thus the penalized sum has one
global minimum with positive definite Hessian. At
\((y,\ldots,y)\) equal power maximizes the channel rate, providing
the saddle and the fading-only converse as in Section~\ref{sec:long}.
For the finite-ratio claim compare with independent complex Gaussian
blocks. At given diagonal gains \(y_j\), Hadamard's determinant
inequality makes the minimum of \(J_{\Id}(X)\) equal to
\(\sum_j j(y_j)\), attained by the diagonal matrix. Dividing the
gamma-model budget by \(m\) gives the same coding and converse
problems as for these Gaussian blocks at exponent \(r/m\) and ratio
\(\lambda/m\). In the converse condition on a ball of gains.
At equal power the auxiliary gains and variances are identical,
so Jensen's inequality bounds all total-energy profiles by the
equal-power rate. The two bounds therefore match for large
\(\lambda\) by Theorem~\ref{thm:exact}. Substituting the Gaussian
multiplier \(\gamma m\) and the ratio \(\lambda/m\) into
\eqref{eq:loss} gives \eqref{eq:nak}.

\section{When is equal power optimal?}\label{sec:equal}
For \(\eta>0\) put \(C_\eta=(C^{-1}+\eta\Id)^{-1}\).
If \(\lambda_1,\ldots,\lambda_K\) are the eigenvalues of \(C\), counted
with multiplicity, the cost of this matrix is
\begin{equation}
 r(\eta)=\sum_{i=1}^K\left\{\log(1+\eta\lambda_i)
                      -\frac{\eta\lambda_i}{1+\eta\lambda_i}\right\}.   \label{eq:eta}
\end{equation}
The function increases continuously from zero to infinity. Its derivative
is \(\sum_i\eta\lambda_i^2/(1+\eta\lambda_i)^2\).

\begin{theorem}[Equal-power criterion]\label{thm:equal}
At reliability \(r=r(\eta)\), equal power attains \(R_\infty(r)\) if and
only if all diagonal entries of \(C_\eta\) are equal. The criterion is
independent of \(\rho>0\). Equal power is optimal at every \(r>0\)
if and only if every spectral projector of \(C\) has constant diagonal.
Equivalently, each power \(C^k\), \(0\le k<m\), has constant diagonal,
where \(m\) is the number of distinct eigenvalues.

If the last condition fails, let \(\ell\ge2\) be the first power with
unequal diagonal entries. There are at most \(m-\ell-1\) positive
reliability levels at which equal power is optimal.
\end{theorem}
\begin{proof}
At equal power every coordinate is active, and \eqref{eq:gradient} is
constant precisely when all adverse variances \(x_j\) coincide.
Equation \eqref{eq:worst} then gives a scalar precision increment
\(\eta\Id\), whose cost fixes \(\eta\) by \eqref{eq:eta}.
Conversely, if \(C_\eta\) has diagonal \(x\mathbf1\), take
\(\gamma=\rho/[K\eta(1+\rho x)]\). Then \(C_\eta\) satisfies
Lemma~\ref{lem:inner} at equal power, and \eqref{eq:gradient} is
constant. Strict concavity proves optimality.

Write \(C=\sum_\lambda\lambda P_\lambda\) over its distinct eigenvalues.
The difference between coordinates \(i,j\) of \(\diag C_\eta\) equals
\[
 f_{ij}(\eta)=\sum_\lambda\frac{w_\lambda}{\eta+1/\lambda},
 \qquad w_\lambda=(P_\lambda)_{ii}-(P_\lambda)_{jj}.
\]
If it vanishes for all \(\eta>0\), its simple-pole residues vanish.
Conversely, equal projector diagonals make every difference vanish.
The Vandermonde system for the distinct eigenvalues relates the
projector diagonals to those of \(C^k\), \(0\le k<m\).
For a pair first separated at power \(\ell\), the series of
\(C_\eta=C-\eta C^2+\eta^2 C^3-\cdots\) about zero gives a zero
of order \(\ell-1\). The common-denominator numerator of \(f_{ij}\)
has degree at most \(m-2\) because \(\sum_\lambda w_\lambda=0\).
After removing the zero at the origin, at most \(m-\ell-1\)
positive roots remain. A level at which all coordinates agree must
be one of them.
\end{proof}

The matrix identities relating equal diagonal powers and local spectral
weights also appear in the theory of walk-regular graphs \cite{Godsil}.
For graph adjacency matrices, Deniskin and Benzi
\cite[Sec.~2.4, Ths.~2.14--2.15]{Deniskin} bound the number of
parameters at which two distinct diagonal resolvent functions agree.
For the covariance
matrix here, the first power with unequal diagonals fixes the
multiplicity at zero and sharpens the count of positive roots.
By \eqref{eq:water} and \eqref{eq:eta}, equal power is optimal at the
specified reliability level precisely when all the diagonal entries
of \(C_\eta\) agree at the corresponding \(\eta\).

\subsection{A finite window of a stationary channel}
Suppose \(C_{ij}=a^{|i-j|}\), \(0<a<1\). Every block has the same
one-dimensional marginal as it would have in an infinite stationary
sequence. A finite codeword can distinguish its edges.
\begin{corollary}\label{cor:AR}
For \(K\ge3\) and every \(r>0\), equal power is strictly suboptimal for
the long-block rate \(R_\infty(r)\).
For \(K=3\) the optimal allocation has equal powers on the two outer
blocks, each strictly larger than the power on the middle block.
For \(K=2\) equal power is optimal at every \(r\).
\end{corollary}
\begin{proof}
Put \(u=a^2\), \(A=1+(1-u)\eta>1\). The tridiagonal matrix
\(B=(1-u)(C^{-1}+\eta\Id)\) has endpoint diagonals \(A\), interior
diagonals \(A+u\), and off-diagonals \(-a\), as direct multiplication
by \(C\) verifies. Let \(D_0=1,D_1=A\) and let \(D_j\) be the
determinant of \(j<K\) consecutive rows and columns ending at the
right endpoint. Expansion gives
\(D_j=(A+u)D_{j-1}-uD_{j-2}\) and hence \(D_j>D_{j-1}\).
The first and second diagonal cofactors differ by
\(D_{K-1}-A D_{K-2}=u(D_{K-2}-D_{K-3})>0\).
Thus \((C_\eta)_{11}>(C_\eta)_{22}\) for every \(\eta>0\),
and Theorem~\ref{thm:equal} applies.

For three blocks, reversal symmetry and uniqueness give
\(p^*=(e,3-2e,e)\). The optimizer is continuous for \(r>0\):
sublevel sets vary continuously there, since mixing any admissible
\(X\) with \(C\) makes its cost strictly smaller than \(r\).
Compactness and uniqueness then give continuity, first of \(F_r\)
and then of its maximizer. By Proposition~\ref{prop:expansion} below,
\(e>1\) for small positive \(r\), because
\((C^2)_{22}-(C^2)_{11}=a^2-a^4>0\).
If \(e\) subsequently reached one, equal power would be optimal
at that level, contrary to the preceding result.
For two blocks the interchange symmetry and uniqueness give equal power.
\end{proof}

\subsection{The gain at small reliability exponents}
In this subsection \(K,C,\rho\) are fixed and \(r\downarrow0\).
Let \(v_j=(C^2)_{jj}=\sum_i|C_{ij}|^2\), \(T_2=\tr C^2\), and
\(\bar v=T_2/K\). The correlation of a block with the rest of the
codeword enters the allocation through \(v_j\).
\begin{proposition}\label{prop:expansion}
The maximizing allocation and its rate gain satisfy
\begin{align}
 p^*_j(r)&=1-\frac{\sqrt{2r}}{\rho\sqrt{T_2}}\,(v_j-\bar v)+O(r),
                                                                         \label{eq:expalloc}\\
 R_\infty(r)-F_r(\mathbf1)
 &=\frac{r}{K(1+\rho)^2T_2}\sum_j(v_j-\bar v)^2+O(r^{3/2}).            \label{eq:expgain}
\end{align}
\end{proposition}
\begin{proof}
Write \(\epsilon=\sqrt{2r}\) and
\(X=C^{1/2}(\Id+\epsilon Z)C^{1/2}\), with the trace inner product
on Hermitian matrices. Uniformly on bounded \(Z\),
\(J_C(X)/\epsilon^2=\tr Z^2/2+O(\epsilon)\), with the expansion
valid also for derivatives on bounded sets. For small \(r\), all
admissible \(Z\) are bounded by the scalar expansion of
\(y-1-\log y\) about one. The first term of
\([S(p,X)-S(p,C)]/\epsilon\) is \(\tr(B_p Z)\), where
\[
 B_p=C^{1/2}\diag\!\left(\frac{\rho p_j}{K(1+\rho p_j)}\right)C^{1/2}.
\]
The minimizing direction at \(\epsilon=0\) is
\(-B_p/\|B_p\|_F\) on \(\tr Z^2=1\). The constrained stationarity
system of these two rescaled functions is smooth through zero. Its
bordered derivative at the minimizing direction has diagonal block
\(\|B_p\|_F\Id\) on the matrix variable and border \(Z\ne0\);
it is invertible by decomposition into the direction \(Z\) and its
orthogonal complement. The implicit function theorem gives a smooth
minimizer locally in \(\epsilon,p\), for \(p\) near \(\mathbf1\).
It is the actual minimizer: compactness and the unique limiting
direction force every sequence of minimizers into this neighborhood.
In particular at \(p=\mathbf1\),
\(X_r(\mathbf1)=C-\epsilon C^2/\sqrt{T_2}+O(\epsilon^2)\).

Let \(\Pi\) subtract the coordinate mean. At equal power
\eqref{eq:gradient} gives the tangent gradient
\[
 -\frac{\epsilon\rho\,\Pi v}{K(1+\rho)^2\sqrt{T_2}}+O(\epsilon^2).
\]
At zero cost the tangent Hessian equals
\(-\rho^2\Id/[K(1+\rho)^2]\). The smooth expansion above shows that
the Hessian varies by \(O(\epsilon+\|p-\mathbf1\|)\) nearby.
Uniform convergence \(F_r(p)\to S(p,C)\) places the maximizer near
the interior point \(\mathbf1\). Solving the tangent stationarity
equation gives \eqref{eq:expalloc}; Taylor expansion of the value
at this displacement gives \eqref{eq:expgain}.
\end{proof}

\subsection{A reversal of the optimal schedule}
An isolated reliability level in Theorem~\ref{thm:equal} can occur
even if every correlation is positive. For four blocks, take
\begin{equation}
 C=\begin{pmatrix}
 1&3/16&1/2&1/2\\3/16&1&1/2&1/2\\
 1/2&1/2&1&1/16\\1/2&1/2&1/16&1
 \end{pmatrix}.                                                        \label{eq:reversal}
\end{equation}
The two within-pair contrasts have covariance
\(C_-=\diag(13/16,15/16)\). The two normalized pair sums have
covariance \(C_+=\bigl(\begin{smallmatrix}19/16&1\\1&17/16\end{smallmatrix}\bigr)\),
of determinant \(67/256>0\). These independent subspaces verify
positive definiteness. Put \(D_\pm=\det(\Id+\eta C_\pm)\).
Taking the diagonal difference in the two subspaces gives
\[
 (C_\eta)_{11}-(C_\eta)_{33}
   =\frac{\eta(\eta-1)}{32D_+D_-}.
\]
Indeed the difference of the two diagonal entries in
\((C_+^{-1}+\eta\Id)^{-1}\) is \((1/8)/D_+\), while the contrast
subspace contributes \(-(1/8)/D_-\); each enters with weight \(1/2\).
At \(\eta=1\) the common diagonal is \(387/899\) and
\begin{equation}
 r_*=\frac{2048}{899}-4+2\log(899/256).                               \label{eq:crossr}
\end{equation}
At every other positive reliability level, equal power is strictly
suboptimal. For small \(r\), the first pair receives less power than
the second by \eqref{eq:expalloc}, since \(v_1-v_3=1/32\).
For large \(r\) the order is reversed. To verify the latter direction,
evaluate the inner problem at uniform power and set \(t=\rho/(K\gamma)\).
As \(r\to\infty\), \(t\to\infty\) by Lemma~\ref{lem:inner}.
Here \(X=(C^{-1}+t\diag\{(1+\rho x_j)^{-1}\})^{-1}\preceq C\),
so \(x_j=O(t^{-1})\). Expanding the inverse first to order
\(t^{-1}\), then to order \(t^{-2}\), gives
\[
 x_j=t^{-1}+[\rho-(C^{-1})_{jj}]t^{-2}+O(t^{-3}).
\]
The same pair decomposition yields
\((C^{-1})_{11}-(C^{-1})_{33}=-2048/13065\), hence \(x_1>x_3\)
for all sufficiently large \(r\). At uniform power this gives a
positive derivative for moving power from the second pair to the
first. The outer objective on the pair-symmetric simplex is strictly
concave. Its maximizer is continuous, and by \eqref{eq:crossr}
it can pass through equal power only at \(r_*\). Thus the first
pair receives less power for every \(0<r<r_*\) and more for every
\(r>r_*\), for every \(\rho>0\).

\subsection{A block can drop out and return}\label{sec:off}
The set of active blocks need not be monotone in the reliability
requirement. Let \(K=3,\ \rho=1/10\), and take
\begin{equation}
 C=\begin{pmatrix}
 1&17/20&1/2\\17/20&1&17/20\\1/2&17/20&1
 \end{pmatrix}.                                                        \label{eq:off}
\end{equation}
The leading principal minors are positive and \(\det C=11/400\).
Set \(p=(3/2,0,3/2)\) and
\(X=(C^{-1}+\diag(1,0,1))^{-1}\).
Inverting gives \(x_1=x_3=7/15,\ x_2=211/500\), and
\[
 J_C(X)=\log(15/4)-14/15 .
\]
At this exponent \(\gamma=5/107\) satisfies \eqref{eq:worst}.
The active marginal rates multiplied by \(K/\rho\) equal \(140/321\);
the inactive marginal is \(211/500<140/321\). These strict inequalities
and \eqref{eq:water} prove that the middle block receives zero power.
The two exponents at which the middle block changes status have a
simple characterization. Replace the two precision increments by
\(t>0\). With \(D(t)=(1+3t/2)(1+t/2)\), inversion gives
\begin{align}
 x_e(t)&=\frac{1+3t/4}{D(t)},\qquad
 x_c(t)=\frac{1+11t/200}{1+3t/2},\notag\\
 r(t)&=\log D(t)-2t x_e(t),                                            \label{eq:offdiag}
\end{align}
where \(e,c\) denote an edge and the center. The two eigenvalues of
the edge covariance are \(1/2,3/2\); differentiating their two
contributions as in \eqref{eq:eta} shows that \(r(t)\) increases from
zero to infinity. For every \(t\) the inner equations hold with
\(\gamma=(3/20)/[3t(1+3x_e(t)/20)]\). The outer inequality is
\(x_e(t)/(1+3x_e(t)/20)\ge x_c(t)\). Clearing positive denominators gives
\[
 -660t^3+4141t^2+1188t-2400\ge0 .
\]
The polynomial is negative at zero, positive at one and negative for
large \(t\). Descartes' rule of signs therefore gives exactly two
positive roots \(t_-<t_+\). The center is inactive precisely on
\([r(t_-),r(t_+)]=[0.227855\ldots,2.142691\ldots]\).
Outside that interval all three blocks carry power. Indeed symmetry
leaves only the alternative of using the center alone. A single
precision increment \(t\) at the center gives
\(x_e-x_c=(1-(17/20)^2)t/(1+t)>0\). The inactive edge would then
have a larger marginal rate than the active center.

At small \(r\), Proposition~\ref{prop:expansion} gives positive power
in all blocks. At sufficiently large \(r\) all blocks are active for
\emph{every} positive definite \(C\). To see this, choose
\(0<\tau<1\) with \(K j(\tau)=r\), so \(\tau C\in\Ec_r\).
Jensen's inequality yields \(R_\infty(r)\le \log(1+\rho\tau)\to0\).
Choose a block \(i\) with \(p_i^*\ge1\). Then \(x_i\to0\) because
\(K R_\infty(r)\ge\log(1+\rho x_i)\), so its marginal rate tends
to zero. If another block \(j\) were inactive on a sequence
\(r\to\infty\), \eqref{eq:worst} and the Schur complement would
give \(x_j\ge1/(C^{-1})_{jj}>0\). Its marginal would exceed the
common active marginal, contradicting optimality. The middle
block in the example therefore returns to the schedule when the
required exponent is sufficiently large.

\subsection{How finite length moves an equal-power point}
The equality criterion also has a finite-length version. Suppose
\(C_\eta\) has constant diagonal \(x\) for some \(\eta>0\).
For a given \(\lambda\) put \(g=\rho x\), \(b=\eta/(\lambda\rho)\),
and assume \(0<b<1\) and \(s=b[1+g(1-b)]/(1-b)\le1\).
Then equal power is exactly optimal under the total energy constraint at
\begin{equation}
 r_\lambda(\eta)=r(\eta)+K\lambda
   \{g b-\log[1+g b(1-b)]\}.                                         \label{eq:crossshift}
\end{equation}
Indeed set \(\xi=1-b,\ w=1+g b(1-b)\) on every block. Then
\(w=(1+s)\xi\), and the precision increment in \eqref{eq:matchX}
equals \(\lambda\rho b=\eta\). Equation \eqref{eq:matchid} gives
\(j(w)+g b^2=g b-\log w\), proving the cost formula.
At equal power coding attains \eqref{eq:exactrate}. The same
auxiliary channel bounds all other profiles: its cost depends only on
total energy and its gains are equal, so equal power maximizes
\eqref{eq:localrate}. The strict-budget limit used in
Theorem~\ref{thm:exact} proves the claim.
Conversely, suppose equal power is optimal and admits a matched
solution. Fix that solution's \(s,u\) in \eqref{eq:ach} while varying
the input powers. This achievable lower bound attains the unrestricted
optimum at the interior equal-power profile. Its partial derivatives,
evaluated at its unique minimizing matrix, are
\(g_j\xi_j/[K(1+g_j\xi_j)]\). They must agree. At fixed \(s\),
\(\xi\) is increasing in \(g\) by \eqref{eq:matchscalar}, so these
equalities force all \(g_j\), and hence all \(X_{jj}\), to coincide.
Equation \eqref{eq:matchX} then has a common precision increment
\(\eta=\lambda\rho(1-\xi)\), and its budget is exactly
\eqref{eq:crossshift}. In particular the criterion is necessary
throughout the range \eqref{eq:threshold}.
For the four-block example \eqref{eq:reversal}, \(\eta=1\) and
\(x=387/899\). The exact equality point therefore moves by
\(4(x/\rho+x^2/2)/\lambda+O(\lambda^{-2})\) from the long-block
value \eqref{eq:crossr}. This is a shift in the reliability exponent
at which equal power is optimal. For \(\rho=1\) the equality point
is \(0.83182\) at \(\lambda=50\), \(0.80074\) at \(\lambda=200\),
and \(0.79030\) in the long-block limit.

\section{How large are the two effects?}\label{sec:numeric}
\subsection{The coding-length correction}
Table~\ref{tab:length} compares the exact rate under the total energy
constraint with the long-block limit. There is no allocation choice
in the one-block channel. In the three-block channel the optimum
leaves the center unused. The scaled loss is
\(D_\lambda=\lambda(R_\infty-\mathcal C_{(L_d)}(r))\).
\begin{table}[htb]
\centering
\caption{Optimal rates and scaled losses as \(d\to\infty\) at fixed
\(\lambda\), in nats per complex use. For three blocks,
\(\rho=0.1\) and \(r=\log(15/4)-14/15\).}\label{tab:length}
\begin{tabular}{rrr@{\qquad}rrr}
\toprule
\multicolumn{3}{c}{One block, \(\rho=3,\ r=.2\)}&
\multicolumn{3}{c}{Three blocks, covariance \eqref{eq:off}}\\
\(\lambda\)&rate&\(D_\lambda\)&\(\lambda\)&rate&\(D_\lambda\)\\
\midrule
4   &.830241&.311614& 64&.040831&.273559\\
8   &.866761&.331072&128&.042870&.286180\\
16  &.886736&.342537&256&.043960&.293245\\
32  &.897244&.348829&512&.044526&.297005\\
64  &.902643&.352136&1024&.044814&.298947\\
128 &.905381&.353834&4096&.045032&.300434\\
\midrule
\(\infty\)&.908145&.355561&\(\infty\)&.045106&.300935\\
\bottomrule
\end{tabular}
\end{table}

Both calculations reduce to one scalar root. Let \(m\) be the
number of active blocks, \(p=K/m\) their common power, \(t\) their
common precision increment and \(x(t)\) their adverse gain.
Set \(b=t/(\lambda\rho p)\) and \(w=1+(t/\lambda)x(t)(1-b)\).
Using \eqref{eq:matchscalar} gives
\(\lambda\rho p b=t\) and \(j(w)+\rho p x(t)b^2=\rho p x(t)b-\log w\).
The trace term in \(J_C\) cancels the linear terms in the budget.
For \(P\) the diagonal projection onto the active blocks, solve
\[
 \log\det(\Id+t C P)-m\lambda\log w=r .
\]
The left side increases for \(0<t<\lambda\rho p\): the log
determinant has derivative \(mx(t)\), whereas
\(\lambda w'=x(t)(1-2b)+t x'(t)(1-b)<x(t)\), and \(w\ge1\).
For \(K=1\), \(x(t)=(1+t)^{-1}\). For \eqref{eq:off} select the two edges and use
\(x(t)=x_e(t)\) from \eqref{eq:offdiag}. In both cases
\(s=w/(1-b)-1\) lies between zero and one in every table row.

For three blocks it remains to rule out any advantage from moving
energy to the center. In the converse choose its noise variance one
and its amplitude reduction \(b_c=b\sqrt{x_e(t)/x_c(t)}<1\).
Then \(x_c(t)b_c^2=x_e(t)b^2\), so every full-budget profile has
the same cost. Write \(A=\rho x_e(t)(1-b)^2/w\).
The marginal gain at an active edge, \(A/(1+3A/2)\), exceeds
the gain \(\rho x_c(t)(1-b_c)^2\) at the unused center. The
difference ranges from \(0.00134\) to \(0.00141\) on the six rows
(rounded). Consequently the water-filling optimum of the auxiliary
channel uses just the two edges and matches the achieved rate.
At \(\lambda=64\) the rate is about \(9.5\%\) below its long-block
limit in this example, versus \(0.61\%\) in the scalar channel.
The two limits of \(D_\lambda\) are comparable; the long-block
rates to which the losses are relative differ by a factor of twenty.

The two channels also illustrate the short-block boundary in
Theorem~\ref{thm:positive}. Positive rate begins at a length ratio
of approximately \(0.14760\) in the one-block case and \(2.92390\) in
the three-block case. To see the second value, set \(a=\lambda\rho/2\).
At \(p=(3/2,0,3/2)\) the determinant in \eqref{eq:positive} is
\(1+3a+27a^2/16\). Its derivative in either edge power exceeds its
derivative in the center power by \(a^2(117/400-99a/1600)\). This
is positive at the root \(a\simeq0.146195\) where the determinant equals
\(e^r\), so the concavity of the log determinant proves optimality.
The sufficient thresholds \(r/c(\rho a_r)\) for matching all
allocations are approximately \(0.94808\) and \(44.20385\),
respectively. For \(\lambda_{\rm pos}(r)<\lambda<r/c(\rho a_r)\)
the channel already supports positive rates, but the sufficient
condition alone does not determine their values.

\subsection{Changing the allocation at the same reliability}
The equal-power criterion involves spectral projectors. To test what a
spectrum alone can predict, compare \(C=1\oplus(0.1\Id_3+
0.9\mathbf1\mathbf1^{\mathsf T})\) with
\(C'=H\diag(2.8,1,0.1,0.1)H^{\mathsf T}\), where \(H\) is a real
Hadamard matrix of order four divided by two. Both covariances have
the displayed eigenvalues and unit diagonals. Only \(C'\) has equal
diagonal weights in every eigenspace. Let \(R_e=F_r(\mathbf1)\).
Table~\ref{tab:iso} compares rate at fixed exponent. Its penultimate
column also compares the two power policies at a common rate within
the first channel. We obtain the exponent \(I\) there by solving
\(F_I(p^*(r))=R_e\); the minimizing cost stays on the boundary and
the positive multiplier makes \(F_I\) strictly decreasing.

\begin{table}[tb]
\centering
\caption{Four blocks with identical spectra. The isolated block in
\(C\) has power \(p_1^*\); each of the others has \((4-p_1^*)/3\).
Rates in both channels are in nats per use.}\label{tab:iso}
\setlength{\tabcolsep}{3pt}
\begin{tabular}{@{}ccrrrrr@{}}
\toprule
&&\multicolumn{4}{c}{One isolated block, \(C\)}
 & Balanced \(C'\)\\
\(\rho\)&\(r\)&\(p_1^*\)&\(R_e\)&\(R_\infty\)&\(I\)&\(R_e=R_\infty\)\\
\midrule
.25&.2&1.6014&.141683&.144479&.217374&.141865\\
.25&.8&1.6989&.084767&.088249&.861597&.084960\\
1&.2&1.3170&.474471&.477590&.206544&.476038\\
1&.8&1.5633&.301394&.309128&.843481&.303574\\
\bottomrule
\end{tabular}
\end{table}

At \(\rho=.25,\ r=.8\) reallocating power raises the rate by
\(4.11\%\) in \(C\) and leaves it unchanged in \(C'\).
The difference can be much smaller in a stationary channel. At
\(\rho=1,r=.2,K=3\), with \(C_{ij}=a^{|i-j|}\), the rate gains are
\(0.044\%,0.026\%,0.010\%\) for \(a=.8,.9,.95\), respectively;
the corresponding middle powers are \(0.9260,0.9414,0.9632\).
There is no positive lower bound on the gain over all \(0<a<1\):
as \(a\downarrow0\), \(C\) tends to the identity, the unique
solutions of \eqref{eq:worst} and the optimal values vary continuously,
and equal power is optimal at \(C=\Id\). To obtain the rates, we
solved \eqref{eq:worst} at fixed \(p\), varied
\(\gamma\) until \(J_C=r\), and equated the marginal rates within
the permutation-symmetric allocation. As a check on the adverse
covariance at each of the equal and optimized allocations, write
\(C=AA^*\) and vary the lower triangular factor \(B\) in
\(X=ABB^*A^*\), with positive diagonal and constraint
\(\|B\|_F^2-2\sum_j\log B_{jj}-K\le r\). For these real \(C\),
uniqueness and conjugation imply that the minimizing covariance is
real. The rates from direct constrained minimization and from
\eqref{eq:worst} differed by less than \(10^{-12}\) for both
allocations in all 45 three-block and eight four-block cases.

\subsection{What a finite receive array can distinguish}
Compare equal power and \(p^*(.2)\) at the \emph{same} threshold
\(R_e=F_{.2}(\mathbf1)\) of each channel. Use three blocks, either
with \(C_{ij}=.9^{|i-j|}\) or with only the first two blocks
correlated, \(C_{12}=C_{21}=.9\). For each channel draw 300,000
independent sets of receive vectors and apply both schedules to the
same draws. The measured event is \(S(p,X_d)<R_e\).
For each comparison, let \(N_{10},N_{01}\) count the realizations
on which only equal power or only the optimized schedule is in
outage. The difference of outage probabilities is the difference of
these two cell probabilities. We bound each with exact binomial
limits \cite{CP}, assigning \(0.05/4\) to each of the four tails,
and subtract endpoints. Coverage is at least \(95\%\) for each
difference. Table~\ref{tab:paired} gives all four comparisons at
\(d=12\), with intervals rounded outwards. To recover the full
allocation from \(p_3^*\), use \(p_1^*=p_3^*\) for the stationary
channel and \(p_1^*=p_2^*\) for the correlated pair, with
\(\sum_j p_j^*=3\) in both cases.

\begin{table}[!htb]
\centering
\caption{Paired fading-outage counts at \(d=12,r=.2\). Both schedules
use the threshold \(R_e\). The last column gives equal-power outage
minus optimized-schedule outage in percentage points.}\label{tab:paired}
\setlength{\tabcolsep}{2.8pt}
\begin{tabular}{@{}lcrrrrrc@{}}
\toprule
Channel&\(\rho\)&\(p_3^*\)&\(R_e\)&Equal&\(p^*(.2)\)&\(N_{10}/N_{01}\)&95\% interval\\
\midrule
Stationary&.25&1.092175&.125441&5554&5531&191/168&$[-.0127,.0281]$\\
Stationary&1&1.029277&.428243&5671&5671&39/39&$[-.0098,.0098]$\\
Correlated pair&.25&1.310613&.143544&5291&4715&1097/521&$[.1499,.2341]$\\
Correlated pair&1&1.180096&.480305&5622&5385&500/263&$[.0498,.1082]$\\
\bottomrule
\end{tabular}
\end{table}
Across \(d=4,12,24\), the lower confidence limit is positive for all
six correlated-pair comparisons and only for \(d=4,\rho=.25\) in the
stationary channel. At \(d=12,\rho=1\), the 78 discordant stationary
draws split evenly.

Write \(a=.9\) and \(y_j=d(X_d)_{jj}\). Given \(y_j=y\), an adjacent correlated
gain \(2y_i/(1-a^2)\) has a noncentral chi-square law with \(2d\)
degrees of freedom and noncentrality \(2a^2y/(1-a^2)\). In the
stationary channel the edge gains are independent given \(y_2\):
their innovations are independent given the center vector and each
conditional law depends only on its norm. For a related factorization,
see \cite{Shi}. The outage condition
\(\prod_{j=1}^3(1+\rho p_j y_j/d)<e^{3R_e}\) gives three successive
upper limits. We integrate over \(y_2,y_1\) and use the conditional
distribution function of \(y_3\) for the stationary channel. For the
correlated pair the corresponding order is \(y_1,y_3,y_2\).
Increasing the Gauss--Legendre order from 96 to 144 on both bounded
intervals changed each of the 24 probabilities by less than \(7\times10^{-16}\).

At \(d=24,\rho=1\) the stationary outage probabilities are
\(0.00126701\) and \(0.00126136\) for equal and optimized power.
The sample difference has the opposite sign, but its interval contains
the integrated difference. For the correlated pair at \(d=24,\rho=.25\)
the probabilities are \(0.00119899\) and \(0.00097796\).

\section{Conclusion}
The adverse covariance determines the long-block power allocation
and the leading rate loss from receiver noise. Even with identical
block marginals, the center and edges can have different optimal
powers, although the rate gain can be small. The critical length
ratio determines when positive rates become possible. Where the
coding and converse bounds differ, the rate remains open in general. Letting
the number of blocks grow would require controlling the adverse
covariance as its dimension increases.

\section*{Acknowledgment}
During manuscript preparation and revision, we used gpt for English language polishing and code organization.


\begin{thebibliography}{99}
\bibitem{Ozarow} L. H. Ozarow, S. Shamai (Shitz), and A. D. Wyner,
``Information theoretic considerations for cellular mobile radio,''
\emph{IEEE Trans. Veh. Technol.}, vol. 43, no. 2, pp. 359--378, 1994.
\bibitem{Foz} M. Fozunbal, S. W. McLaughlin, and R. W. Schafer,
``On performance limits of space-time codes: A sphere-packing bound approach,''
\emph{IEEE Trans. Inf. Theory}, vol. 49, no. 10, pp. 2681--2687, 2003.
\bibitem{Yang} W. Yang, G. Durisi, T. Koch, and Y. Polyanskiy,
``Quasi-static multiple-antenna fading channels at finite blocklength,''
\emph{IEEE Trans. Inf. Theory}, vol. 60, no. 7, pp. 4232--4265, 2014.
\bibitem{Hoydis} J. Hoydis, R. Couillet, and P. Piantanida,
``The second-order coding rate of the MIMO quasi-static Rayleigh fading
channel,'' \emph{IEEE Trans. Inf. Theory}, vol. 61, no. 12, pp. 6591--6622, 2015.
\bibitem{Kara} A. Karadimitrakis, A. L. Moustakas, and R. Couillet,
``Gallager bound for MIMO channels: Large-\(N\) asymptotics,''
\emph{IEEE Trans. Wireless Commun.}, vol. 17, no. 2, pp. 1323--1330, 2018.
\bibitem{Nak} B. Nakibo\u{g}lu, ``The sphere packing bound for memoryless
channels,'' \emph{Problems Inform. Transm.}, vol. 56, no. 3, pp. 201--244, 2020.
\bibitem{Font} J. Font-Segura, A. Martinez, and A. Guill\'en i F\`abregas,
``Asymptotics of the random-coding union bound in quasi-static fading
channels,'' in \emph{Proc. IEEE Information Theory Workshop}, 2016,
pp. 251--254.
\bibitem{PPV} Y. Polyanskiy, H. V. Poor, and S. Verd\'u,
``Channel coding rate in the finite blocklength regime,''
\emph{IEEE Trans. Inf. Theory}, vol. 56, no. 5, pp. 2307--2359, 2010.
\bibitem{PPVparallel} Y. Polyanskiy, H. V. Poor, and S. Verd\'u,
``Dispersion of Gaussian channels,'' in \emph{Proc. IEEE Int. Symp.
Information Theory}, 2009, pp. 2204--2208.
\bibitem{Shin} H. Shin and M. Z. Win, ``Gallager's exponent for MIMO
channels: A reliability-rate tradeoff,'' \emph{IEEE Trans. Commun.},
vol. 57, no. 4, pp. 972--985, 2009.
\bibitem{Kaz} P. Kazakopoulos, P. Mertikopoulos, A. L. Moustakas, and
G. Caire, ``Living at the edge: A large deviations approach to the outage
MIMO capacity,'' \emph{IEEE Trans. Inf. Theory}, vol. 57, no. 4,
pp. 1984--2007, 2011.
\bibitem{Caire} G. Caire, G. Taricco, and E. Biglieri,
``Optimum power control over fading channels,'' \emph{IEEE Trans. Inf.
Theory}, vol. 45, no. 5, pp. 1468--1489, 1999.
\bibitem{Coon} J. P. Coon and R. Cepeda, ``Power loading in parallel
diversity channels based on statistical channel information,''
\emph{IEEE Commun. Lett.}, vol. 16, no. 7, pp. 1060--1063, 2012.
\bibitem{Zhang} W. Zhang and U. Mitra, ``On outage behavior of wideband
slow-fading channels,'' in \emph{Proc. 45th Annu. Allerton Conf.
Communication, Control, and Computing}, 2007.
\bibitem{Abbe} E. Abbe, S.-L. Huang, and E. Telatar, ``Proof of the
outage probability conjecture for MISO channels,'' \emph{IEEE Trans.
Inf. Theory}, vol. 59, no. 5, pp. 2596--2602, 2013.
\bibitem{Bai} B. Bai, W. Chen, K. B. Letaief, and Z. Cao, ``Outage
exponent: A unified performance metric for parallel fading channels,''
\emph{IEEE Trans. Inf. Theory}, vol. 59, no. 3, pp. 1657--1677, 2013.
\bibitem{Loyka} S. Loyka and C. D. Charalambous, ``On the compound
capacity of a class of MIMO channels subject to normed uncertainty,''
\emph{IEEE Trans. Inf. Theory}, vol. 58, no. 4, pp. 2048--2063, 2012.
\bibitem{XuZhaoPaths} H. Xu and H. Zhao, ``Large deviations and conditional
laws for a time-integrated logarithm of averaged Gaussian squares,''
unpublished manuscript, 2026.
\bibitem{XuZhaoStatic} H. Xu and H. Zhao, ``When does the quasi-static
approximation predict MIMO outage?'' manuscript, 2026.
\bibitem{DZ} A. Dembo and O. Zeitouni, \emph{Large Deviations Techniques
and Applications}, 2nd ed. Berlin, Germany: Springer, 1998.
\bibitem{Gallager} R. G. Gallager, \emph{Information Theory and Reliable
Communication}. New York, NY, USA: Wiley, 1968.
\bibitem{CoverThomas} T. M. Cover and J. A. Thomas, \emph{Elements of
Information Theory}, 2nd ed. Hoboken, NJ, USA: Wiley, 2006.
\bibitem{Tarokh} V. Tarokh, N. Seshadri, and A. R. Calderbank,
``Space-time codes for high data rate wireless communication:
Performance criterion and code construction,''
\emph{IEEE Trans. Inf. Theory}, vol. 44, no. 2, pp. 744--765, 1998.
\bibitem{Dog} A. Dogand\v{z}i\'c, ``Chernoff bounds on pairwise error
probabilities of space-time codes,'' \emph{IEEE Trans. Inf. Theory},
vol. 49, no. 5, pp. 1327--1336, 2003.
\bibitem{Shannon} C. E. Shannon, ``Probability of error for optimal
codes in a Gaussian channel,'' \emph{Bell Syst. Tech. J.}, vol. 38,
no. 3, pp. 611--656, 1959.
\bibitem{KL} G. A. Kabatiansky and V. I. Levenshtein,
``On bounds for packings on a sphere and in space,''
\emph{Problems Inform. Transm.}, vol. 14, no. 1, pp. 1--17, 1978.
\bibitem{Godsil} C. D. Godsil and B. D. McKay, ``Feasibility conditions
for the existence of walk-regular graphs,'' \emph{Linear Algebra Appl.},
vol. 30, pp. 51--61, 1980.
\bibitem{Deniskin} N. Deniskin and M. Benzi, ``New results and open
problems on subgraph centrality,'' \emph{J. Combin.}, vol. 14, no. 4,
pp. 425--444, 2023.
\bibitem{CP} C. J. Clopper and E. S. Pearson, ``The use of confidence
or fiducial limits illustrated in the case of the binomial,''
\emph{Biometrika}, vol. 26, no. 4, pp. 404--413, 1934.
\bibitem{Shi} Z. Shi, S. Ma, G. Yang, K.-W. Tam, and M. Xia,
``Asymptotic outage analysis of HARQ-IR over time-correlated Nakagami-\(m\)
fading channels,'' \emph{IEEE Trans. Wireless Commun.}, vol. 16, no. 9,
pp. 6119--6134, 2017.
\end{thebibliography}
\end{document}